\documentclass[11pt,preprint,prd,aps,superscriptaddress,nofootinbib,showpacs,a4paper]{revtex4}

\newcommand{\lapprox}{%
\mathrel{%
\setbox0=\hbox{$<$}\raise0.6ex\copy0\kern-\wd0\lower0.65ex\hbox{$\sim$}}}

\newcommand{\be}{\begin{equation}}
\newcommand{\en}{\end{equation}}

 \usepackage{graphicx}
\usepackage{multirow}
\usepackage{amssymb}
\usepackage{amsmath}
\newcommand{\bea}{\begin{eqnarray}}
\newcommand{\eea}{\end{eqnarray}}

\begin{document}




\title{$K^*$ resonance effects on direct $CP$ violation in $B \to \pi \pi K$}


\author{O.~Leitner}
\affiliation{Laboratoire de Physique Nucl\'eaire et de Hautes \'Energies (IN2P3--CNRS--Universit\'es Paris 6 et 7), Groupe Th\'eorie, \\
                  Universit\'es Pierre et Marie Curie et Paris-Diderot, 4 place Jussieu, 75252 Paris, France}

\author{J.-P. Dedonder}
\affiliation{Laboratoire de Physique Nucl\'eaire et de Hautes \'Energies (IN2P3--CNRS--Universit\'es Paris 6 et 7), Groupe Th\'eorie, \\
                  Universit\'es Pierre et Marie Curie et Paris-Diderot, 4 place Jussieu, 75252 Paris, France}

\author{ B.~Loiseau}
\affiliation{Laboratoire de Physique Nucl\'eaire et de Hautes \'Energies (IN2P3--CNRS--Universit\'es Paris 6 et 7), Groupe Th\'eorie, \\
                  Universit\'es Pierre et Marie Curie et Paris-Diderot, 4 place Jussieu, 75252 Paris, France}

\author{R. Kami\'nski}
\affiliation{Division of Theoretical Physics, The Henryk Niewodnicza\'nski Institute
 of Nuclear Physics, \\
Polish Academy of Sciences, 31-342  Krak\'ow, Poland}

\date{\today }

\begin{abstract}
$B^{\pm} \to \pi^+ \pi^- K^{\pm}$ and $\bar{B}^0 \to \pi^+ \pi^- \bar{K}^0$ decay channels are analyzed within 
the QCD factorization scheme where final state interactions before and after hadronization are included.
The $K^{*}(892)$ and $K^{*}_0(1430)$ resonance effects are taken into account using the presently known $\pi K$ strange vector and scalar form factors.
The weak decay amplitudes, which are calculated at leading power in $\Lambda_{QCD}/m_b$ and at the next-to-leading order in the strong coupling constant, include
the hard scattering and annihilation contributions.
The end point divergences of these weak final state interactions are controlled by two complex parameters determined through a fit to the available effective mass and helicity angle distribution, $CP$ asymmetry and  $K^{*}(892)$ branching ratio data.
The predicted  $K^{*}_0(1430)$ branching ratios  and the calculated direct $CP$ violation asymmetries are compared to the  Belle and BABAR Collaboration data. 
\end{abstract}





\maketitle

\section{Introduction}\label{intro}

In the Standard Model, it is known  that $CP$ violation is mainly predicted in weak decays because of  
the weak phase of the Cabibbo-Kobayashi-Maskawa (CKM) matrix~\cite{CKM}. In decays of $B$ mesons, through  electroweak 
interaction, one can calculate the matter anti-matter asymmetry;   $W$-boson exchange and  large  
beauty quark mass, allow a systematic perturbative calculation in the QCD factorization formalism (QCDF)~\cite{QCDF} where the final state interactions 
are the main source of uncertainty. It is the combined occurrence of a weak and a strong phase differences 
that lead to the observation of the $CP$ violating asymmetry between the $B \to \pi^{\pm} \pi^{\mp} K$ and $\bar{B} 
\to \pi^{\mp} \pi^{\pm} \bar{K}$ channels. 

Electroweak decays of resonant and non-resonant mesons made of a $q\bar{q}$ pair are well described in  QCDF. 
In this framework, there is no direct three-body factorization scheme that efficiently describes a three-body decay, hence a quasi two-body 
state has first to be built up.
Here, one of the two mesons assumed to be a $K^*$ resonance can decay via a strong decay mechanism to a $(\pi K)$ state. In 
Ref.~\cite{Bppk}, the authors attempt to reproduce the $\pi K$ effective mass and helicity angle distributions.
In that calculation, the weak amplitude relies on  effective QCD coefficients describing the leading order contribution as well as the 
 vertex and penguin corrections at the order of $\Lambda_{QCD}/m_b$. The $K^*$ resonances decaying into $\pi K$ are then modeled by 
the scalar and vector  form  factors~\cite{Moussallam:2007qc} that correspond to the strong final state interactions after hadronization. Additional 
phenomenological amplitudes, represented by four complex free
parameters and added to the QCD penguin amplitude, are fitted to mainly reproduce the $B \to K^*(892) \pi$ branching ratio and the $CP$ asymmetry of the recent Belle and 
BABAR collaboration data. Furthermore they also  predict the $B \to K^*(1430) \pi$ branching
 ratio. Altogether, one obtains a fair description of the data for these three-body $B$ decays. 

In the present work, one explicitly 
includes the hard scattering and annihilation corrections at the order of $\Lambda_{QCD}/m_b$. 
These weak final state interactions based on phenomenological assumptions are controlled by the endpoint divergences related to 
the asymptotic wave functions. This approach reduces thus the number of free parameters to only two complex ones. 

In Sec.~\ref{3bodyamp}, we derive the three-body decay amplitudes for the $B \to \pi \pi K$ 
processes within the QCDF framework introducing quasi two-body states. Sections~\ref{QCDF892} and \ref{QCDF1430}  provide all the details for the weak decay amplitudes calculated at next-to-leading order in the strong coupling constant 
and
in the perturbative expansion of the short distance interaction for $B \to \pi K^{* }(892)$ and 
$B \to \pi K_0^{* }(1430)$.  
Section~\ref{Input} lists  all the numerical parameters employed and in Section~\ref{Resper} a discussion follows the presentation of  the significant  results on branching ratios and asymmetries.
Finally, Section~\ref{Conclusion} concludes with a summary of our work and some outlook.

\section{Three-body decay amplitude}\label{3bodyamp}

To analyze the  $B \to \pi \pi K$ decay amplitude, one first evaluates the matrix element 
$\langle  \pi M_2 \,\vert \, \mathcal{H}_{eff} \, \vert \, B\,\rangle$ within the factorization hypothesis,
\begin{equation}
\langle  \pi M_2 \,\vert \, \mathcal{H}_{eff} \, \vert \, B\,\rangle \propto 
 \langle  M_2 \,\vert \, \bar s \, \gamma_\nu (1-\gamma_5) d  \, \vert \, 0 \rangle 
\langle  \pi  \,\vert \,  \bar u \, \gamma^\nu (1-\gamma^5) b \, \vert \, B\, \rangle,
\end{equation}
with $M_2$ being either the vector  $K^*(892)$ or scalar $K_0^{* }(1430)$ resonance,  $\mathcal{H}_{eff}$ is the standard effective 
Hamiltonian for $B$ decay (see Ref.~\cite{Bppk}). The vector $K^*(892)$ and the scalar $K_0^{* }(1430)$ resonances are assumed to be 
$(\pi K)$ quasi bound states 
in $P$ and $S$ waves, respectively. Thus, one writes~\cite{Bppk},
\begin{equation}\label{B3c}
\langle  \pi \pi K \,\vert \, \mathcal{H}_{eff} \, \vert \, B\,\rangle \propto \langle  (\pi K)_{S,P} \,\vert \, \bar s \, \gamma_\nu (1-\gamma_5) d  \, \vert \, 0 \rangle 
\langle  \pi  \,\vert \,  \bar u \, \gamma^\nu (1-\gamma^5) b \, \vert \, B\, \rangle,
\end{equation}
where $\langle  (\pi K)_{S,P} \,\vert \, \bar s \, \gamma_\nu (1-\gamma_5) d  \, \vert \, 0 \rangle$ is expressed as
\begin{equation}
\label{Ktopiff}
\langle \pi(p_{\pi}) K(p_{K})\vert\bar s\gamma_\nu(1-\gamma_5) d\vert 0\rangle =
\Biggl[
(p_{K}-p_{\pi})_\nu   -\dfrac{m_{K}^2-m_{\pi}^2}{q^2}q_\nu
\Biggr]
f_1^{\pi K}(q^2) 
+\dfrac{m_{K}^2-m_{\pi}^2}{q^2}q_\nu f_0^{\pi K}(q^2).
\end{equation}
In Eq.~(\ref{Ktopiff}), $q^2$ with $q=p_{K}+p_{\pi}$ is the invariant $\pi K$ mass squared, $m_K$ and $m_{\pi}$ denote the kaon and pion masses, respectively. The  vector $f_{1}^{\pi K}(q^2)$ and scalar 
$f_{0}^{\pi K}(q^2)$  form factors are describing the final 
state interaction after hadronization. From semileptonic decays like $\tau \to K \pi \nu_{\tau}$ and $K \to \pi l \nu_l$, one can extract informations on these $K\pi$ scalar
and vector form factors~\cite{Moussallam:2007qc}. Analyticity, unitarity, QCD counting asymptotic rules allow one to relate scalar and vector form factors to the 
$K^*_0(1430) \to \pi K $ and $K^*(892) \to \pi K$ scattering amplitudes in the elastic and inelastic domains. All the details can be found in Refs.~\cite{Bppk} 
and~\cite{Moussallam:2007qc}. The full amplitude for each wave is given by
\begin{equation}
\label{abppk}
\mathcal{A}_3(B \to \pi \pi K)=\mathcal{A}(B \to \pi M_2) \times\Gamma(M_2 \to K \pi).
\end{equation}
For the $K^*(892)$, the vertex function $\Gamma(K^*(892) \to K \pi)$ associated with the $B \to \pi K^* \to \pi \pi K$ decay 
is written as
\begin{equation}\label{vertexv}
 \Gamma(K^*(892) \to K \pi)=\frac{2}{q f_{K^*}} \, \frac{{\mathbf p}_{\pi^+} \cdot {\mathbf p}_{\pi^-}}
{\epsilon_{K^{* }(892)}^*\!\cdot p_B} f_1^{\pi K}(q^2),
\end{equation}
where $f_{K^*}$ is the $K^*$ 
decay constant and $\epsilon_{K^{* }(892)}^* \!\cdot p_B = (m_B/2 q) \; \lambda^{1/4}\left(m_B^2,q^2,m_{\pi}^2\right)$,  $p_B$ and $m_B$ denoting the $B$ four momentum and mass, respectively.
 In Eq.~(\ref{vertexv}),  $\lambda(x,y,z)=(x+y-z)^2-4 x y$ and the moduli of the $\pi^{\pm}$ momenta are, 
\begin{equation}\label{ppi+}
\left|{\mathbf p}_{\pi^+}\right|=
\dfrac{1}{2 q} 
\sqrt{
\Bigl[
q^2-\left(m_{K}+m_{\pi}\right)^2
\Bigr]
\Bigl[q^2-\left(m_{K}-m_{\pi}\right)^2\Bigr]
},
\end{equation}
and,
\begin{equation}\label{ppi-}
\vert{\mathbf p}_{\pi^-}\vert=
\dfrac{1}{2q} 
\sqrt{
\Bigl[
m_B^2-\left(q+m_{\pi}\right)^2
\Bigr]
\Bigl[m_B^2-\left(q-m_{\pi}\right)^2\Bigr]
}.
\end{equation} 
The vertex function $\Gamma(K_0^{* }(1430)\to K \pi)$ is 
\begin{equation}
\label{fks}
\Gamma(K^*_0(1430) \to K \pi)= \frac{1}{f_{K^*_0}}\frac{m_K^2-m_{\pi}^2}{q^2}f_0^{\pi K}(q^2),
\end{equation}
where $f_{K^*_0}$ denotes the $K^*_0$ decay constant. Following  closely~\cite{QCDF}, one derives the QCDF decay amplitudes where the short and long distance 
contributions are factorized in the approximation of a quasi two-body state,  $\pi K^*(892)$ or $\pi K^*_0(1430)$. 

The amplitude 
$B^- \to \pi^-  \bar{K}^{*0}(892)$ is
\begin{multline}\label{ampkv1}
   {\cal A}(B^- \to \pi^-  \bar{K}^{*0})
   = 
    \sum_{q=u,c} \lambda_q^{(s)} 
   \Bigg\{ A_{\pi  K^*} \biggl[ \delta_{qu}\, \beta_2(\mu)  + a_4^q(\mu)  
 + r_{\chi}^{K^*}(\mu) a_6^q(\mu)\\  -\frac12 \Bigl(a_{10}^q(\mu)
 + r_{\chi}^{K^*}(\mu) a_8^q(\mu) \Bigr)
   + \beta_3(\mu)  +\beta_{3,{\rm EW}}(\mu) \biggr]_{\pi K^*}
   \Bigg\},
\end{multline}
and the $\bar{B}^0 \to \pi^+  \bar{K}^{*-}(892)$ amplitude,
\begin{multline}\label{ampkv2}
   {\cal A}(\bar{B}^0 \to \pi^+ \bar{K}^{*-})
   = 
   \sum_{q=u,c} \lambda_q^{(s)} 
   \Bigg\{A_{\pi  K^*} \biggl[ \delta_{qu}\, a_1^q(\mu)
   + a_4^q(\mu)  +  r_{\chi}^{K^*}(\mu) a_6^q(\mu) \\+a_{10}^q(\mu) + r_{\chi}^{K^*}(\mu) a_8^q(\mu)  +\beta_3(\mu)
  - \frac12  \beta_{3,{\rm EW}}(\mu)\biggr]_{\pi K^*}
   \Bigg\},
\end{multline}
where the coefficients $a_n^q(\mu)$ and $\beta_n(\mu)$ are given in Eqs.~(\ref{a}) and~(\ref{beta}). The  $\lambda_q^{(s)}$  are product of CKM matrix elements, the
 $r_{\chi}^{M_2}(\mu)$  the chiral coefficients and $\mu$ is the scale.

 The  
$B^- \to \pi^-  \bar{K}^{*0}_0(1430)$,  amplitude reads,
\begin{multline}\label{ampks1}
   {\cal A}(B^- \to \pi^- \bar{K}^{*0}_0)
   =
 \sum_{q=u,c} \lambda_q^{(s)} 
   \Bigg\{ A_{\pi K_0^{*}} \biggl[ \delta_{qu}\, \beta_2(\mu)   + a_4^q(\mu)  
 - r_{\chi}^{K^*_0}(\mu) a_6^q(\mu) \\-\frac12 \Biggl(a_{10}^q(\mu) 
- r_{\chi}^{K^*_0}(\mu) a_8^q(\mu) \Biggr) + \beta_3(\mu)+\beta_{3,{\rm EW}}(\mu) \biggr]_{\pi K_0^{*}}
   \Bigg\},
\end{multline}
while  the $\bar{B}^0 \to \pi^+ K_0^{* -}(1430)$ amplitude is
\begin{multline}\label{ampks2}
   {\cal A}(\bar{B}^0 \to \pi^+ K_0^{* -})
   = \sum_{q=u,c} \lambda_q^{(s)} 
   \Bigg\{A_{\pi K_0^{*}} \biggl[ 
   \delta_{qu}\, a_1^q(\mu) 
   + a_4^q(\mu)  -  r_{\chi}^{K^*_0}(\mu) a_6^q(\mu) \\+a_{10}^q(\mu) 
     -
    r_{\chi}^{K^*_0}(\mu) a_8^q(\mu)
    + \beta_3(\mu)- \frac12  \beta_{3,{\rm EW}}(\mu)\biggr]_{\pi K_0^{*}}
   \Bigg\}.
\end{multline}
 The chiral coefficients, $r_{\chi}^{K^*}(\mu)$ and $r_{\chi}^{K^*_0}(\mu)$, will be given in Eqs.~(\ref{rchikv}) and~(\ref{rchiks}). 

For the $K^*(892)$ resonance, the pseudoscalar-vector factor $A_{\pi K^{*}}$ in Eqs.~(\ref{ampkv1}) and~(\ref{ampkv2}) reads,
\begin{equation}\label{apv}
A_{\pi K^{*}}=
-i\,\frac{G_F}{\sqrt2} \, 2 q\,\epsilon_{K^{* }(892)}^*\!\cdot p_B\, 
F_0^{B\to \pi}(q^2) f_{K^*},
\end{equation}
with the Fermi constant $G_F=1.16 \times 10^{-5} \rm{GeV}^{-2}$ and where the weak transition form factor $F_0^{B\to \pi}(q^2)$ will be given in Sec.~\ref{Input}.
For the $K^*_0(1430)$ scalar resonance, the pseudoscalar-scalar factor $A_{\pi K^{*}_0}$ in Eqs.~(\ref{ampks1}) and~(\ref{ampks2}), is
\begin{equation}\label{aps}
A_{\pi K^{*}_0}=i\,\frac{G_F}{\sqrt2} (m_B^2-m_{\pi}^2) F_0^{B\to \pi}(q^2) f_{K^*_0}. 
\end{equation}
 
In Eqs.~(\ref{ampkv1})-(\ref{ampks2}), the CKM matrix elements are,
 \begin{align}\label{ckm}
\lambda_u^{(s)} = V_{ub}V_{us}^*&= A \lambda^3 \left(\rho - i \eta \right)\lambda, \nonumber \\
\lambda_c^{(s)} = V_{cb}V_{cs}^*&= A \lambda^2 \left(1-\frac{\lambda^2}{2}\right),
\end{align}
where following Ref.~\cite{PDG} the Wolfenstein parameters are, $A=0.814$, $\rho=~0.1385$, $\eta=0.358$ and $\lambda=0.2257$. 

Since one
assumes the dominance of the $K^*(892)$ and $K^*_0(1430)$  resonances in the description of the $\pi K$ channel, the full amplitude 
$\mathcal{A}_3(B \to \pi \pi K)$ is built up on the $P$ and $S$ waves so that the differential effective mass branching fraction is~\cite{Bppk},
\begin{equation}\label{dB-}
\dfrac{d\mathcal{B}(B \to \pi \pi K)}{dq}=\dfrac{1}{\Gamma_{B}}
\dfrac{q \ \vert{\mathbf p}_{\pi^+}\vert\ \vert{\mathbf p}_{\pi^-}\vert}{4(2\pi)^3 m_B^3} 
\Biggl(
\Bigl|\mathcal{A}_3(B \to \pi (\pi K)_S)\Bigr|^2+ 
\dfrac{1}{3} \Bigl|\mathcal{A}_3(B \to \pi (\pi K)_P)\Bigr|^2
\Biggr),
\end{equation}
where   $\Gamma_{B}=1/\tau_B$ is the $B$-decay width. The usual $CP$ 
violating asymmetry parameter is
\begin{equation}
\label{ACP}
\mathbb{A}_{CP}(B \to \pi \pi K)=\frac{\mathcal{B}(B \to \pi \pi K)-\mathcal{B}(\bar{B} \to 
\bar{\pi} \bar{\pi} \bar{K})}{\mathcal{B}(B \to \pi \pi K)+\mathcal{B}(\bar{B} \to \bar{\pi} \bar{\pi} \bar{K})}.
\end{equation}

In Eqs.~(\ref{ampkv1})-(\ref{ampks2}), the $a_n^q(\mu)$, involving the Wilson coefficients $C_n(\mu)$, are
\begin{multline}\label{a}
  a_n^q(\mu) = \left [C_n(\mu) + \frac{C_{n\pm 1}(\mu)}{N_c} \right ] N_n(M_2)  
   +  P_n^q(M_2) \\
   + \frac{C_f\ }{4\pi N_c}\, \Biggl[ \alpha_s(\mu)C_{n\pm 1}(\mu)  V_n(M_2)
+ \frac{4 \pi^2
    \alpha_s(\mu/2)}{N_c} C_{n\pm 1}(\mu/2) H_n(\pi M_2) \Biggr], 
\end{multline}
with $n \in \{1,10\}$, and the scale is $\mu=m_b$,  $m_b$ being the $b$ quark mass. In Eq.~(\ref{a}), the color number is $N_c=3$ and $C_f=(N_c^2-1)/2 N_c=4/3$.
The upper (lower) signs in $C_{n\pm 1}(\mu)$ apply when $n$ is odd (even) and
\begin{equation}\label{n}
   N_n (M_2) = 
     \begin{cases}
       0, & n\in\{6,8\},\,\text{and}\, M_2 \equiv K^{*}(892),\\
       1, & \text{else.}
     \end{cases}
\end{equation}

The Wilson coefficients, $C_n(\mu)$, computed in the Naive Dimension Regularization (NDR) scheme~\cite{QCDF}, are taken at the scale
 $m_b$ for the vertex, $V_n(M_2)$, and penguin $P_n^q(M_2)$, corrections which involve only the $b$-quark,
whereas  the annihilation, $\beta_n(\pi M_2)$ and hard scattering, $H_n(M_2)$, contributions are evaluated at the 
scale $m_b/2$ since they involve the spectator quark. The strong coupling constants are $\alpha_s(m_b)=0.224$ and $\alpha_s(m_b/2)=0.286$~\cite{PDG}.

The annihilation term, $\beta_n(\mu)$, is given by
\begin{equation}\label{beta}
\left [\beta_n(\mu)\right ]_{\pi M_2}=
\frac{\left [b_n(\mu)\right ]_{\pi M_2} B_{\pi M_2}}{A_{\pi M_2}},
\end{equation}
where the factor, $B_{\pi M_2}$, is the product of $G_F$ by the $B,\ \pi$ and $M_2$ decay constants, 
\begin{equation}\label{Bpksv}
  B_{\pi M_2} =  \mp i\,\frac{G_F}{\sqrt{2}}\,f_{B} f_{\pi} f_{M_2},
\end{equation}
with the upper sign if $M_2 \equiv K^{* }(892)$ and the lower sign otherwise. 
In 
Eqs.~(\ref{ampkv1})-(\ref{ampks2}), the tree annihilation 
component (at $\mu=m_b/2$) is (the upper-scripts $I$ and $F$ denote initial and final states),
\begin{equation}\label{bt}
  \left [b_2(\mu)\right ]_{\pi M_2} = \frac{C_f}{N_c^2}\,C_2(\mu) A_1^I(\pi M_2), 
\end{equation}
while the penguin annihilation terms (at $\mu=m_b/2$)  are
\begin{multline}
    \left [ b_3(\mu)\right ]_{\pi M_2} = \frac{C_f}{N_c^2} \Bigg[ C_3(\mu) A_1^I(\pi M_2) + C_5(\mu) \Bigl(A_3^I(\pi M_2)  
      +A_3^F(\pi M_2)\Bigr)+ N_c C_6(\mu) A_3^F(\pi M_2) \Bigg],\nonumber 
\end{multline}
\vskip -0.5cm
\begin{multline}\label{bp}
   \left [ b_{3,\rm EW}(\mu)\right ]_{\pi M_2} = \frac{C_f}{N_c^2} \Bigg[ C_9(\mu) A_1^I(\pi M_2) 
    + C_7(\mu) \Bigl(A_3^I(\pi M_2) +A_3^F(\pi M_2) \Bigr) 
    + N_c C_8(\mu) A_3^F(\pi M_2) \Bigg], 
\end{multline}
where the amplitudes $A_j^{I,F}(\pi M_2)$ are given in Eqs.~(\ref{annikv}) for the $P$-wave and~(\ref{anniks}) for the $S$-wave.  

\section{QCDF corrections for $B \to \pi K^{* }(892) $}\label{QCDF892}

The pion in the final state $\pi K^{* }(892)$ is created from the transition $B \to \pi$ while the $K^{* }(892)$ is 
created from the vacuum; this mechanism is due to the structure of the four-quark operators in the heavy quark effective theory as 
well as the conservation of the flavor quantum numbers.  Following Ref.~\cite{QCDF}, we only give the QCD corrections that appear in 
$\mathcal{A}(B \to \pi K^{* }(892))$.

Since the coefficients in the Gegenbauer expansion of the light cone distribution amplitudes (LCDA) are known with large uncertainties~\cite{QCDF}, one here 
limits oneself to leading terms in this expansion for the $\pi$ and $K^{* }(892)$. The leading twist-2 distribution 
amplitude is $\Phi(x)=6 x (1-x)$ and the twist-3 two particle distribution is $\varphi(x)=1$ and  $\varphi(x)=3(2 x-1)$ for both $\pi$ and $K^{* }(892)$.

The chiral coefficient for the vector meson $K^{* }(892)$, given at the scale $\mu$ is defined as 

\begin{equation}\label{rchikv}
r_{\chi}^{K^*}(\mu)= \frac{2 \sqrt{q^2}}{m_b(\mu)}\frac{f_{K^*}^{\perp}}{f_{K^*}},
\end{equation}
where $f_{K^*}^{\perp}$ is the transverse decay constant and where one has introduced the running meson mass square, replacing $m_{K^*(892)}^2$ by $m_{\pi K}^2=q^2$. For a pion, the chiral coefficient reads
 \begin{equation}\label{rchip}
r_{\chi}^{\pi}(\mu)= \frac{m_{\pi}^2}{m_b(\mu)m_u(\mu)},
\end{equation}
with the $u$-quark mass $m_u$.

\subsection{Penguin contributions}\label{Peng892}

The penguin contributions, $P_n^q(K^{* }(892))$, with the values $n=4,~6,~8,~10$, required in Eqs.~(\ref{ampkv1}) and~(\ref{ampkv2}), are as follows,
\begin{multline}\label{p4kv}
   P_4^q(K^{* }(892)) = \frac{C_f \ \alpha_s(\mu)}{4\pi N_c}\,\Biggl\{
    C_1(\mu) \Biggl[ \frac43\ln\frac{m_b}{\mu}
    + \frac23 - G_{K^{* }(892)}(s_q) \Biggr]\\
    + C_3(\mu) \;\biggl[ \frac83\ln\frac{m_b}{\mu} + \frac43 
    - G_{K^{* }(892)}(0) - G_{K^{* }(892)}(1) \biggr] 
    + \Bigl(C_4(\mu)+C_6(\mu)\Bigr) 
    \times \Biggl[ \frac{4n_F}{3}\ln\frac{m_b}{\mu}\\
    - (n_F-2)\,G_{K^{* }(892)}(0) -  G_{K^{* }(892)}(s_c)  - G_{K^{* }(892)}(1) \Biggr] 
   - 2 C_{8g}^{\rm eff}(\mu) \int_0^1 \frac{dx}{1-x}\,
    \Phi_{K^{* }(892)}(x) \Biggr\},
\end{multline}
with
\begin{equation}
\int_0^1 \frac{dx}{1-x}\Phi_{K^{* }(892)}(x)=3,
\end{equation}
and $C_{8g}^{\rm eff}(\mu)$ related to the $Q_{8g}$ chromomagnetic dipole operator. Furthermore,
\begin{multline}\label{p6kv}
   P_6^q(K^{* }(892)) = - \frac{C_f \ \alpha_s(\mu)}{4\pi N_c}\,\Biggl\{
    C_1(\mu)\,\hat G_{K^{*}(892)}(s_q) + 
     C_3(\mu)\,\Biggl[ \hat G_{K^{* }(892)}(0)  
      + \hat G_{K^{* 0}(892)}(1) \Biggr] \\
   + \Bigl(C_4(\mu)+C_6(\mu)\Bigr)  \Biggl[ (n_F-2)\,\hat G_{K^{*}(892)}(0)
    + \hat G_{K^{*}(892)}(s_c) + \hat G_{K^{*}(892)}(1) \Biggr] \Biggr\},
\end{multline}
\begin{equation}\label{p8kv}
   P_8^q(K^{* }(892)) =  - \frac{\alpha_e}{9\pi N_c}\,\Bigl(C_1(\mu)+N_c C_2(\mu)\Bigr)\,
   \hat G_{K^{* }(892)}(s_q),
\end{equation}
where $\alpha_e=1/129$ is the electromagnetic coupling constant. Finally, 
\begin{multline}\label{p10kv}
   P_{10}^q(K^{*}(892)) = \frac{\alpha_e}{9\pi N_c}\,\Biggl\{
   \Bigl(C_1(\mu)+N_c C_2(\mu)\Bigr) \Biggl[ \frac{4}{3}\ln\frac{m_b}{\mu}  + \frac23  
   - G_{K^{*}(892)}(s_q) \Biggr]\\
   - 3 C_{7\gamma}^{\rm eff}(\mu) \int_0^1 \frac{dx}{1-x}\,\Phi_{K^{*}(892)}(x)
   \Biggr\}. 
\end{multline}
In these equations,  $\mu=m_b$ and the number of active flavors is $n_F=5$.  In Eq.~(\ref{p10kv}), $C_{7\gamma}^{\rm eff}(\mu)$  is 
related to the $Q_{7\gamma}$ electromagnetic dipole operator. The gluon kernel contributions are  

\begin{equation}\label{gkv}
  G_{K^{* }(892)}(s_q) = \left\{
    \begin{array}{ll}
       \displaystyle
      \frac53 + \frac{2 i \pi}{3}, \qquad \hskip -0.5em s_q =0,&
       \\[1em]
       \displaystyle
       \frac{85}{3} -6 \sqrt{3} \pi + \frac{4 \pi^2}{9}\ , \qquad \hskip -0.5em s_q=1, &
       \\[1.em]
       \displaystyle
          \frac53 - \frac23 {\rm ln} s_c + \frac{32}{3} s_c + 16 s_c^2 
-\displaystyle \frac23 \sqrt{1-4 s_c} 
         \Bigl[1+ 2 s_c +24 s_c^2   \Bigr]
\displaystyle \Bigl [ 2 \arctan {\rm h}\left(\sqrt{1-4 s_c}\right) \Bigr .
\\ 
\Bigl . -i\pi  \Bigr ]      
\displaystyle 
+ 12 s_c^2 \left(1- \frac43 s_c  \right) 
 \Bigl[ 2 \arctan \!{\rm h}\left(\sqrt{1-4 s_c}\right)-i\pi  \Bigr]^2,
     \hskip -0.5em \qquad s_q=s_c,&
    \end{array}
  \right.
\end{equation}
and

 \begin{equation}\label{gckv}
  \hat{G}_{K^{* }(892)}(s_q) =\left\{
    \begin{array}{ll}
       \displaystyle
      1, \qquad \hskip -0.5em s_q =0,&
       \\[1em]
       \displaystyle
       -35 + 4 \sqrt{3} \pi + \frac{4 \pi^2}{3}, \qquad 
\hskip -0.5em s_q=1, &
       \\[1.em]
       \displaystyle
       -12 \; s_c^2 \; \Bigl [2 \arctan \!{\rm h}\left(\sqrt{1-4 s_c}\right)-i\pi\Bigr]^2  -36\ s_c       
       \\ [1.em] \displaystyle
 + 12 \; \sqrt{1-4 s_c}\; s_c \; 
\displaystyle \Bigl[2 \arctan \!{\rm h}\left(\sqrt{1-4 s_c}\right)-i\pi\Bigr]+1,  \qquad \hskip -0.5em s_q=s_c.&
    \end{array}
  \right.
\end{equation}
In Eqs.~(\ref{gkv}) and~(\ref{gckv}), $s_q$ is  defined as $(m_q/m_b)^2$ so that $s_q=0$ for $q=u,d$, $s_q=1$ for $q=b$ and  $s_q=s_c$ for $q=c$.

\subsection{Vertex contributions}\label{Ver892}

In the $B \to \pi K^{* }(892)$ transition, the electroweak vertex, $V_n(K^{*}(892))$,  receives $\alpha_s(\mu)$ corrections to all $a_n^q(\mu)$ in the 
amplitude $\mathcal{A}(B \to \pi K^{*}(892))$, 
 \begin{equation}\label{vkv}
  V_n(K^{*}(892)) = \left\{
    \begin{array}{ll}
       \displaystyle
       12 \; {\rm ln} \left (\frac{m_b}{\mu}\right)  - 3 i \pi - \frac{37}{2},
\hskip -0.5em \qquad 
n \in \{1,4,10\}, &
       \\[1.em]
       \displaystyle
        9 - 6 i \pi\ ,  \hskip -0.5em \qquad 
n \in \{6,8\}.&
    \end{array}
  \right.
\end{equation}

\subsection{Hard scattering contributions}\label{Har892}

Evaluated at the scale $\mu=m_b/2$, the hard scattering correction can be written as,

\begin{equation}
\label{hnpM_2}
\displaystyle
H_n(\pi M_2)=\frac{B_{\pi M_2}}{A_{\pi M_2}} \tilde{H}_n(\pi M_2),
\end{equation} 
where for the vector resonance, $M_2 \equiv K^{*}(892)$, $A_{\pi K^{*}(892)}$ and $B_{\pi K^{*}(892)}$ are
 defined by Eqs.~(\ref{apv}) and~(\ref{Bpksv}), respectively. One has
\begin{equation}\label{hkv}
  \tilde{H}_n(\pi K^{* }(892)) = \left\{
    \begin{array}{ll}
       \displaystyle
     3 \frac{m_B}{\lambda_B} \left [ r_\chi^{\pi}(\mu) \; X_H + 3  \right],
\qquad 
\hskip -0.5em n \in \{1,4,10\}, &
       \\[1em]
       \displaystyle
        0\ ,  \hskip -0.5em \qquad 
n \in \{6,8\},&
    \end{array}
  \right.
\end{equation}
where $\lambda_B=0.3$ GeV is a hadronic parameter of the order of $\Lambda_{QCD}$~\cite{Beneke:2001ev}. In Eq.~(\ref{hkv}),  $r_\chi^{\pi}(\mu)$ is  given 
by Eq.~(\ref{rchip}) and $X_H$ represents the end point divergence related to the soft-gluon interaction with the spectator quark. Its expression  will 
be given in Eq.~(\ref{defxah}) in Sec.~\ref{Input}. 

\subsection{Annihilation contributions}\label{Ann892}

The annihilation amplitudes cannot be derived from the QCDF approach so that they are model-dependent involving also a divergence
parameterized by $X_A$ (Eq.~(\ref{defxah})). Based on Ref.~\cite{QCDF}, the expressions for $A_j^I(\pi K^{* }(892))$ and 
$A_j^F(\pi K^{* }(892))$, for $j=1$ and $3$, are,

\begin{equation}\label{annikv}
\begin{aligned}
 &  A_1^I(\pi K^{* }(892)) \approx 6\pi\alpha_s(\mu)\; 
\Biggl[\,
    3\,\bigg( X_A - 4 + \frac{\pi^2}{3} \bigg)
    + r_\chi^{K^{* }}(\mu) r_\chi^{\pi}(\mu) \;(X_A^2-2 X_A) \Biggr] , \\
 &  A_3^I(\pi K^{* }(892)) \approx 6\pi\alpha_s(\mu) \;\Biggl[ -3 r_\chi^{K^{*}}(\mu)\; 
\bigg( X_A^2 - 2 X_A - \frac{\pi^2}{3} + 4 \bigg)
     + r_\chi^{\pi}(\mu) \;\bigg( X_A^2 - 2 X_A + \frac{\pi^2}{3} \bigg) 
    \Biggr] , \\[0.3cm] &  
A_3^F(\pi K^{* }(892)) \approx -6\pi\alpha_s(\mu)\; \Biggl[\, 3 r_\chi^{K^{* }}(\mu)\, 
(2 X_A-1) (2-X_A) -  r_\chi^{\pi}(\mu)\,(2 X_A^2 - X_A) \Biggr],
\end{aligned}
\end{equation}
with $\mu=m_b/2$.

\section{QCDF corrections for  $B \to \pi K_0^{* }(1430)$}\label{QCDF1430}

We now turn to the $B \to \pi K_0^{* }(1430)$ transition for which the $\alpha_s(\mu)$ corrections are all included. Here again, only the 
first non-vanishing leading term  in the LCDA  of the 
$K_0^{* }(1430)$ are retained:

\begin{equation}
\Phi_{K_0^{* }(1430)}(x)= 6x (1-x) \left[1+3 B_1^{K_0^{* }(1430)}(\mu)\; (2 x -1)    \right],
\end{equation}
where $B_1^{K_0^{* }(1430)}(\mu=m_b)= 5.26$, and $B_1^{K_0^{* }(1430)}(\mu=m_b/2)= 0.39$ are the first non-vanishing Gegenbauer moment (for neutral scalar) 
evaluated at two different mass scales. The asymptotic form of the LCDA for the pion is
\begin{equation}
\Phi_{\pi}(x)= 6x (1-x).
\end{equation}
The twist-3 two particle distributions are 
\begin{equation}
\varphi_{K_0^{* }(1430)}(x) = 1 \;\; {\rm and} \;\;\varphi_{\pi}(x)  =1. 
\end{equation}

Similarly to the $B \to \pi K^{* }(892)$ decay channel, the $B \to \pi K_0^{* }(1430)$ decay amplitude is factorized out into a product of a transition form
 factor $B \to \pi$ times a $K_0^{* }(1430)$ decay constant as shown in Eq.~(2) of Ref.~\cite{QCDF}.

The $K_0^{* }(1430)$ chiral coefficient is given by:
\begin{equation}\label{rchiks}
r_{\chi}^{K_0^{* }(1430)}(\mu) =  \frac{2 q^2}{m_b(\mu) \left(m_s(\mu)-m_u(\mu) \right)},
\end{equation}
where $m_s$ is the strange quark mass. In Eq.~(\ref{rchiks}), as has been done for the $K^*(892)$ meson (see Eq.~(\ref{rchikv}), one has introduced the running meson 
mass square for the $K_0^{* }(1430)$ replacing $m_{K^*_0(1430)}^2$ by $m_{\pi K}^2=q^2$.

\subsection{Penguin contributions}\label{Peng1430}

From Ref.~\cite{QCDF}, one can obtain all the penguin corrections $P_n^q(K_0^{* }(1430))$,   (with $n=4,6,8,10$) for the $B$ to pseudoscalar-scalar transition. 
One has,
\begin{multline}\label{p4ks}
   P_4^q(K_0^{* }(1430)) = \frac{C_f \ \alpha_s(\mu)}{4\pi N_c}\,\Biggl\{
    C_1(\mu) \Biggl[ \frac43\ln\frac{m_b}{\mu}
    + \frac23   - G_{K_0^{* }(1430)}(s_q) \Biggr] \\
     + C_3(\mu) \Biggl[ \frac83\ln\frac{m_b}{\mu} + \frac43 
    - G_{K_0^{* }(1430)}(0) - G_{K_0^{* }(1430)}(1) \Biggr] 
    + \Bigl(C_4(\mu)+C_6(\mu)\Bigr)
    \Biggl[ \frac{4n_F}{3}\ln\frac{m_b}{\mu}
    - (n_F-2)\\ \times G_{K_0^{* }(1430)}(0) - G_{K_0^{* }(1430)}(s_c) - G_{K_0^{* }(1430)}(1) \Biggr] 
   - 2 C_{8g}^{\rm eff}(\mu) \int_0^1 \frac{dx}{1-x}\,
    \Phi_{K_0^{* }(1430)}(x) \Biggr\},
\end{multline}
with 
\begin{equation}
\int_0^1 \frac{dx}{1-x}\,\Phi_{K_0^{* }(1430)}(x)=3 B_1^{K_0^{* }}(\mu) + 3.
\end{equation}
Moreover,
\begin{multline}\label{p6ks}
   P_6^q(K_0^{* }(1430)) = \frac{C_f \ \alpha_s(\mu)}{4\pi N_c}\,\Biggl\{
    C_1(\mu) \Biggl[ \frac43\ln\frac{m_b}{\mu}
    + \frac23  - \hat G_{K_0^{* }(1430)}(s_q) \Biggr]
   + 
C_3(\mu) \Biggl[ \frac83\ln\frac{m_b}{\mu} + \frac43 \\
    - \hat G_{K_0^{* }(1430)}(0)  - \hat G_{K_0^{* }(1430)}(1) \Biggr] 
  + \Bigl(C_4(\mu)+C_6(\mu)\Bigr)
    \Bigl[ \frac{4n_F}{3}\ln\frac{m_b}{\mu} \\
    - (n_F-2) \hat G_{K_0^{* }(1430)}(0) -  \hat G_{K_0^{* }(1430)}(s_c) 
        - \hat G_{K_0^{* }(1430)}(1)
    \Biggr] - 2 C_{8g}^{\rm eff}(\mu) \Biggr\}, 
\end{multline}
\begin{equation}\label{p8ks}
   P_8^q(K_0^{* }(1430)) = \frac{\alpha_e}{9\pi N_c}\,\Biggl\{
   \Bigl(C_1(\mu)+N_c C_2(\mu)\Bigr) \Biggl[ \frac{4}{3}\ln\frac{m_b}{\mu} + \frac23  
   - \hat G_{K_0^{* }(1430)}(s_q) \Biggr] - 3 C_{7\gamma}^{\rm eff}(\mu) \Biggr\},
\end{equation}
and
\begin{multline}\label{p10ks}
   P_{10}^q(K_0^{* }(1430)) = \frac{\alpha_e}{9\pi N_c}\,\Biggl\{
   \Bigl(C_1(\mu)+N_c C_2(\mu)\Bigr)   \Biggl[ \frac{4}{3}\ln\frac{m_b}{\mu}  + \frac23  
   - G_{K_0^{* }(1430)}(s_q) \Biggr] \\
   - 3 C_{7\gamma}^{\rm eff}(\mu) \int_0^1 \frac{dx}{1-x}\,\Phi_{K_0^{* }(1430)}(x)
   \Biggr\}. 
\end{multline}
Comparing Eq.~(\ref{p4kv}) with  Eq.~(\ref{p4ks}) and Eq.~(\ref{p10kv}) with Eq.~(\ref{p10ks}), one can see that the formal structures of $P_4^q(K_0^{* }(1430))$ and $P_{10}^q(K_0^{* }(1430))$ in terms of $C_n(\mu)$, of gluon kernel functions,  $G_{M_2}(s_q)$ and  of LCDA, $\Phi_{M_2} (x)$,  where $M_2$ is now $K_0^{* }(1430)$ instead of $K^{* }(892)$, are identical to those of $P_4^q(K^{* }(892))$ and $P_{10}^q(K^{* }(892))$, respectively. 
The gluon kernel functions, 
 entering in Eqs.~(\ref{p4ks})-(\ref{p10ks}), are

 \begin{equation}\label{gks}
  G_{K_0^{* }(1430)}(s_q) =\left\{
    \begin{array}{ll}
       \displaystyle
      \frac53 + \frac{2 i \pi}{3}+ \frac{B_1^{K_0^{* }}(\mu)}{2}, \qquad \hskip -0.5em s_q =0, &
       \\[1em]
       \displaystyle
       \frac{85}{3} -6 \sqrt{3} \pi + \frac{4 \pi^2}{9} 
       \displaystyle
-\left[\frac{155}{2}-36 \sqrt{3}\pi + 12 \pi^2\right]B_1^{K_0^{* }}(\mu),  \qquad   \hskip -0.5em s_q=1, &
       \\[1em]
       \displaystyle
          \frac53 - \frac23 {\ln} s_c + \frac{B_1^{K_0^{* }}(\mu)}{2} + \frac{4}{3}\left [8+9 B_1^{K_0^{* }}(\mu)\right ]s_c +   2
        \displaystyle 
\left [8+63  B_1^{K_0^{* }}(\mu)\right ] s_c^2  \\  [1em]

       -306 B_1^{K_0^{* }}(\mu) s_c^3 - \frac23 \sqrt{1-4 s_c}   
        \displaystyle 
\left[1+ 2 s_c +6(4+ 27 B_1^{K_0^{* }}(\mu)) s_c^2  - 324 B_1^{K_0^{* }}(\mu) s_c^3 \right]
        \\  [1em] 
       \displaystyle 
       \times \Bigl[ 2 \arctan \!{\rm h}\left(\sqrt{1-4 s_c}\right)-i\pi  \Bigr] + 12 s_c^2\    
       \displaystyle 
\left[1  + 3 B_1^{K_0^{* }}(\mu) - 
       \frac43 (1+9 B_1^{K_0^{* }}(\mu))s_c \right.\\  [1em]
 \left.+ 18 B_1^{K_0^{* }}(\mu) s_c^2        \right] 
       \displaystyle 
\Bigl[ 2 \arctan \!{\rm h}\left(\sqrt{1-4 s_c}\right)-i\pi  \Bigr]^2, \hskip -0.5em \qquad s_q=s_c,&
    \end{array}
  \right.
\end{equation}
and

  \begin{equation}\label{gcks}
   \hat{G}_{K_0^{* }(1430)}(s_q) =\left\{
    \begin{array}{ll}
        \displaystyle
      \frac{16}{9}+ \frac{2\pi}{3}i, \qquad \hskip -0.5em s_q =0,&
       \\[1em]
       \displaystyle
       \frac{-32}{9} + \frac{2\pi}{\sqrt{3}}\ , \qquad \hskip -0.5em s_q=1, &
       \\[1.em]
       \displaystyle
       \frac{16}{9} (1-3 s_c)-\frac{2}{3} \Bigl[  {\rm ln} s_c + (1-4 s_c)^{3/2}
\displaystyle  \left[2 \arctan \!{\rm h}\left(\sqrt{1-4 s_c}\right)-i\pi\right] \Bigr]\ , \hskip -0.5em \qquad s_q=s_c.&
    \end{array}
  \right.
\end{equation}

\subsection{Vertex contributions}\label{Ver1430}

The relevant vertex corrections, $V_n(K_0^{* }(1430))$, for $n \in \{1,4,6,8,10\}$ are the following, 

\begin{equation}\label{vks}
  V_n(K_0^{* }(1430)) = \left\{
    \begin{array}{ll}
       \displaystyle
       12 \; {\ln} \left (\frac{m_b}{\mu}\right)  - 3 i \pi - \frac{37}{2}
      \displaystyle
      +\frac{1}{2}\left (11- 6 i \pi \right) B_1^{K_0^{* }}(\mu),
       \hskip -0.5em \qquad n \in \{1,4,10\}, &
       \\[1.em]
       \displaystyle
        - 6,  \hskip -0.5em \qquad n \in \{6,8\},&
    \end{array}
  \right.
\end{equation}
with $\mu=m_b$.

\subsection{Hard scattering contributions}\label{Har1430}

From gluon exchange between the scalar $K_0^{* }(1430)$ and the spectator $u$ quark one  derives, at  $\mu=m_b/2$, the hard scattering 
corrections.
 One writes it as in Eq.~(\ref{hnpM_2}) with $M_2 \equiv K_0^{*}(1430)$ and  $A_{\pi K_0^{*}(1430)}$ and
 $B_{\pi K_0^{*}(1430)}$ defined by Eqs.~(\ref{aps}) and~(\ref{Bpksv}), respectively. Here

\begin{equation}\label{hks}
  \tilde{H}_n(\pi K_0^{* }(1430)) = \left\{
    \begin{array}{ll}
       \displaystyle
      \frac{3 m_B}{\lambda_B} \left[3 (B_1^{K_0^{* }}(\mu)+1) - 
       \displaystyle
     r_\chi^{\pi}(\mu) X_H (B_1^{K_0^{* }}(\mu)-1) \right],
\qquad \hskip -0.5em n \in \{1.4,10\}, &
       \\[1.em]
       \displaystyle
        0,  \hskip -0.5em \qquad n \in \{6,8\}.&
    \end{array}
  \right.
\end{equation}
 As for the $K^*(892)$ (Sec.~\ref{Har892}) the endpoint divergence is modeled by $X_H$.
\subsection{Annihilation contributions}\label{Ann1430}
The weak initial and final annihilation amplitudes, $A_j^I(\pi K_0^{* }(1430))$ and $A_j^F(\pi K_0^{* }(1430))$ (with $j=1$ and $3$) at $\mu=m_b/2$ are
calculated starting from 
Ref.~\cite{QCDF} for $B \to \pi K_0^{* }(1430)$: 
\begin{equation}\label{anniks}
  \begin{split}
&  A_1^I(\pi K_0^{* }(1430)) \approx     2 \pi \alpha_s(\mu) \;  \Biggl( 9 B_1^{K_0^{* }}(\mu) (3 X_A+4-\pi^2) 
 - r_\chi^{\pi}(\mu) \;r_\chi^{K_0^{* }}(\mu)\; X_A^2 \Biggr), \\
&  A_3^I(\pi K_0^{* }(1430)) \approx 6 \pi \alpha_s(\mu) \; \Biggl\{ 3 r_\chi^{\pi}(\mu) B_1^{K_0^{* }}(\mu) \Biggl( X_A^2-4 X_A 
+4+\frac{\pi^2}{3} \Biggr)+ r_\chi^{K_0^{* }}(\mu) \Biggl( X_A^2-2 X_A+\frac{\pi^2}{3} \Biggr) \Biggl\}, \\
&  A_3^F(\pi K_0^{* }(1430)) \approx 
  6 \pi \alpha_s(\mu) X_A  \Biggl\{ r_\chi^{\pi}(\mu) B_1^{K_0^{* }}(\mu)(6 X_A
-11) - r_\chi^{K_0^{* }}(\mu) (2 X_A-1)   \Biggr\},
  \end{split}
\end{equation}
with $X_A$ an endpoint divergence (Eq.~(\ref{defxah})). These 
amplitudes will be then implemented in the $b_n(\pi K_0^{* }(1430))$ given in Eqs.~(\ref{bt}) and~(\ref{bp}).

{\renewcommand\baselinestretch{1.10}
\begin{table*}[t]
\begin{center}
\caption{Real and imaginary parts of the leading order (LO), vertex, penguin and hard-scattering contributions to the short distance amplitude, $a_{n}^q(\mu)$, for  $P$ and $S$ waves. The scale $\mu=m_b$ except for the hard scattering where $\mu=m_b/2$.\label{ai}}
\begin{tabular}{|cccccc|}
\hline 
           &                               &                                        & $P$-wave                     &                           &  \\
\hline\hline 
           &   LO                       &  Vertex               & Penguin  & Hard scattering           & Total\\
 \hline \hline
 $a_1^u(\mu)$   & \multirow{2}{*}{$(1.018;0)$}   & \multirow{2}{*}{$(0.028;0.014)$}       & \multirow{2}{*}{$(0;0)$}     & \multirow{2}{*}{$(-0.246;0.317)$} & \multirow{2}{*}{$(0.800;0.331)$}\\
 $a_1^c(\mu)$   &                               &                                        &                      &                                 &  \\
\hline  
 $a_4^u(\mu)$   & \multirow{2}{*}{$(-0.031;0)$}  & \multirow{2}{*}{$(-0.002;-0.001)$}         & $(0.003;-0.014)$      & \multirow{2}{*}{$(0.018;-0.023)$} & $(-0.012;-0.038)$  \\
$a_4^c(\mu)$    &                               &                                        & $(-0.002;-0.005)$      &                                 & $(-0.017;-0.029)$ \\
\hline  
 $a_6^u(\mu)$   & \multirow{2}{*}{ $(0;0)$} & \multirow{2}{*}{$(0.0006;-0.001)$}        & $(-0.007;-0.0009)$      & \multirow{2}{*}{$(0;0)$}          & $(-0.006;-0.002)$  \\
$a_6^c(\mu)$    &                               &                                        & $(0.001;0.011)$       &                                 & $(0.002;0.010)$  \\
\hline  
$a_8^u(\mu)$    & \multirow{2}{*}{$(0;0)$}   & \multirow{2}{*}{$(-0.6;1.3) \times 10^{-5}$ }       & $(-4.7;0)\times 10^{-5}$         & \multirow{2}{*}{$(0;0)$}          & $(-5.3;1.3)\times 10^{-5}$  \\
$a_8^c(\mu)$    &                               &                                        & $(-0.3;6.4)\times 10^{-5}$       &                                 & $(-0.9;7.7) \times 10^{-5}$  \\
\hline  
$a_{10}^u(\mu)$ & \multirow{2}{*}{$(-0.0014;0)$}  & \multirow{2}{*}{$(0.0014;0.0007$)}         & $(0.0002;-0.0001)$       & \multirow{2}{*}{$(-0.013;0.016)$} & \multirow{2}{*}{$(-0.012;0.017)$} \\
$a_{10}^c(\mu)$ &                               &                                        & $(0.0002;-0.0001)$       &                                 &  \\
 \hline  
\hline
           &                               &                                        & $S$-wave                     &                           &  \\
 \hline \hline
 $a_1^u(\mu)$   & \multirow{2}{*}{$(1.018;0)$}   & \multirow{2}{*}{$(-0.016;0.089)$}      & \multirow{2}{*}{(0;0)}     & \multirow{2}{*}{$(-0.151;0.184)$} & \multirow{2}{*}{$(0.851;0.273)$}\\
 $a_1^c(\mu)$   &                               &                                        &                            &                                 &  \\
\hline  
 $a_4^u(\mu)$   & \multirow{2}{*}{$(-0.031;0)$}  & \multirow{2}{*}{$(0.001;-0.007)$}        & $(0.023;-0.017)$           & \multirow{2}{*}{$(0.011;-0.014)$} & $(0.004;-0.037)$  \\
$a_4^c(\mu)$    &                               &                                        & $(0.039;0.036)$            &                                 & $(0.021;0.016)$ \\
\hline  
 $a_6^u(\mu)$   & \multirow{2}{*}{ $(-0.039;0)$} & \multirow{2}{*}{$(-0.0004;0)$}             & $(-0.003;-0.014)$            & \multirow{2}{*}{$(0;0)$}          & $(-0.042;-0.014)$  \\
$a_6^c(\mu)$    &                               &                                        & $(-0.006;-0.004)$            &                                 & (-0.045;-0.004)  \\
\hline  
$a_8^u(\mu)$    & \multirow{2}{*}{$(44;0)\times 10^{-5}$}   & \multirow{2}{*}{$(0.4;0)\times 10^{-5}$ }             & $(4;-10)\times 10^{-5}$             & \multirow{2}{*}{(0;0)}          & $(48;-10)\times 10^{-5}$  \\
$a_8^c(\mu)$    &                               &                                        & $(2;-5)\times 10^{-5}$             &                                 & $(46;-5) \times 10^{-5}$  \\
\hline  
$a_{10}^u(\mu)$ & \multirow{2}{*}{$(-0.0014;0)$}  & \multirow{2}{*}{$(-0.0008;0.005)$}        & $(0.0015;-0.0001)$             & \multirow{2}{*}{$(-0.008;0.009)$} & $(-0.009;0.014)$ \\
$a_{10}^c(\mu)$ &                               &                                        & $(0.0016;0.0002)$              &                                 & $(-0.009;0.014)$ \\
 \hline  
\end{tabular}
\end{center}
\end{table*}}

\section{Input}\label{Input}

\subsection{Numerical parameters}\label{Numpar}

In this Section, one summarizes all the values of the parameters required for performing numerical applications. From 
Ref.~\cite{PDG}, the meson masses in GeV are
\begin{equation}\label{mmesons}
m_{B}= 5.300,\  m_{\pi}=0.139,\ m_{K}=0.493,\ 
m_{K^*}=0.892,\  m_{K^*_0}=1.430,\ m_{B^*}= 5.320.
\end{equation}
The running quark masses (at $m_b=4.2$ GeV) in GeV are
\begin{equation}\label{mquarks}
m_b=4.2,\ m_c=1.3,\ m_s=0.070,\ m_{u,d}=0.003,
\end{equation}
whereas at  $m_b/2$, one has in GeV~\cite{Cheng:2005nb}, 
\begin{equation}\label{mquarks2}
m_b=4.95,\ m_c=1.51,\ m_s=0.090,\ m_{u,d}=0.005.
\end{equation}
The meson decay constants in MeV are

\begin{equation}
\label{decay }
f_{B}=180\pm 40~\mbox{\cite{Beneke:2001ev}}
,\ f_{K^*}=218\pm 4~\mbox{\cite{QCDF}}
,\ f_{\pi}=130\pm 0.2~\mbox{\cite{PDG}}
,\ f^{\perp}_{K^*}=175\pm 25~\mbox{\cite{QCDF}}.
\end{equation}
The scalar meson decay constant $f_{K^*_0}$, 
which appears in Eqs.~(\ref{fks}), (\ref{aps}) and (\ref{Bpksv}), does not, in fact, enter in our calculation as it cancels out in $\mathcal{A}_3(B \to \pi \pi K)$ [Eq.~(\ref{abppk})], in 
$\beta_n(\pi M_2)$ [Eq.~(\ref{beta})] and in $H_n(\pi M_2)$ [Eq.~(\ref{hnpM_2})].

The $B^{\pm}$ and $B^0$ mean lives, entering in Eq.~(\ref{dB-}), are~\cite{PDG} $\tau_{B^{\pm}}=(1.638 \pm 0.011) \times 10^{-12}$ s and 
$\tau_{B^0}=(1.530 \pm 0.009)\times 10^{-12}$ s, respectively. 

For the Wilson coefficients, $C_n(\mu)$, we take, at both scales $\mu = m_b$ and $m_b/2$, the next-to-leading order logarithmic approximation values as given in Table 1 of Ref.~\cite{Beneke:2001ev}.
Using Eqs.~(\ref{a}) and (\ref{n}), one obtains, at the scale $\mu = m_b$, the universal leading order (LO)  $a_n^q(\mu)$ values presented in the first column of Table~\ref{ai}.

{\renewcommand\baselinestretch{1.10}
\begin{table*}[pt]
\begin{center}
\caption{Real and imaginary parts of the annihilation contributions for $P$ and $S$waves. Here $\mu = m_b/2$.\label{betan}}
\begin{tabular}{|ccc|}
 \hline 
                                &     $P$-wave               & $S$-wave  \\
\hline \hline 
 $\beta_{2}(\pi M_2)$     &   $(0.006;0.0007)$         & $(0.031;0.013)$ \\
\hline  
 $\beta_{3}(\pi M_2)$                    &   $(-0.024;-0.011$)        & $(0.094;0.051)$ \\
\hline  
 $\beta_{3,EW}(\pi M_2)$&   $(0.025;0.005)\times 10^{-2}$         & $(-0.009;-0.003)\times 10^{-2}$  \\
\hline  
\end{tabular}
\end{center}
\end{table*}}

\subsection{Model parameters}\label{Modpar}

For the $B \to \pi$ transition form factor, we employ the pole-extrapolation model~\cite{Melikhov:2001zv},
\begin{equation}\label{ffbp}
\displaystyle
F_0^{B\to \pi}(q^2)= \frac{f_0(0)}{\Biggl(1- \sigma_1 \displaystyle
\frac{q^2}{m_{B^*}^{2}} + \sigma_2 \displaystyle
\frac{q^4}{m_{B^*}^4}\Biggr)  },
\end{equation}
at the momentum transfer, $q$. In the transition form factor
model we are using, the numerical parameters are $f_0(0)=0.29$, $\sigma_1=0.76$ and $\sigma_2=0.28$. 

As pointed out in Sec.~\ref{3bodyamp}, we use the  vector $f_{1}^{\pi K}(q^2)$ and scalar 
$f_{0}^{\pi K}(q^2)$  (with $f_K/f_\pi=0.193$) form factors derived in~\cite{Bppk}. 

The hard scattering and annihilation contributions for the $K^{*}(892)$ given in Eqs.~(\ref{hkv}) and~(\ref{annikv}) as well as those for the $K_0^{* }(1430)$ 
given in Eqs.~(\ref{hks}) and~(\ref{anniks}) involve divergences, $X_H$ and $X_A$ which are modeled~\cite{QCDF} as follows,
\begin{equation}\label{defxah}
X_{A,H}= \Bigl( 1+ \rho_{A,H} \exp (i \phi_{A,H})  \Bigr) \ln \frac{m_B}{\lambda_h},
\end{equation}
with, for each $X_{A,H}$, two real parameters $\rho_{A,H} > 0$ and $0 < \phi_{A,H} < 360^o$. One expects  the annihilation and 
hard scattering contributions to be of the order of $\ln (m_B/\lambda_h)$ with $\lambda_h=0.5$ GeV (see Ref.~\cite{Beneke:2001ev}).

\section{Results and Discussion}\label{Resper}

\begin{figure}[ht]
\includegraphics*[angle=90,width=1.\columnwidth]{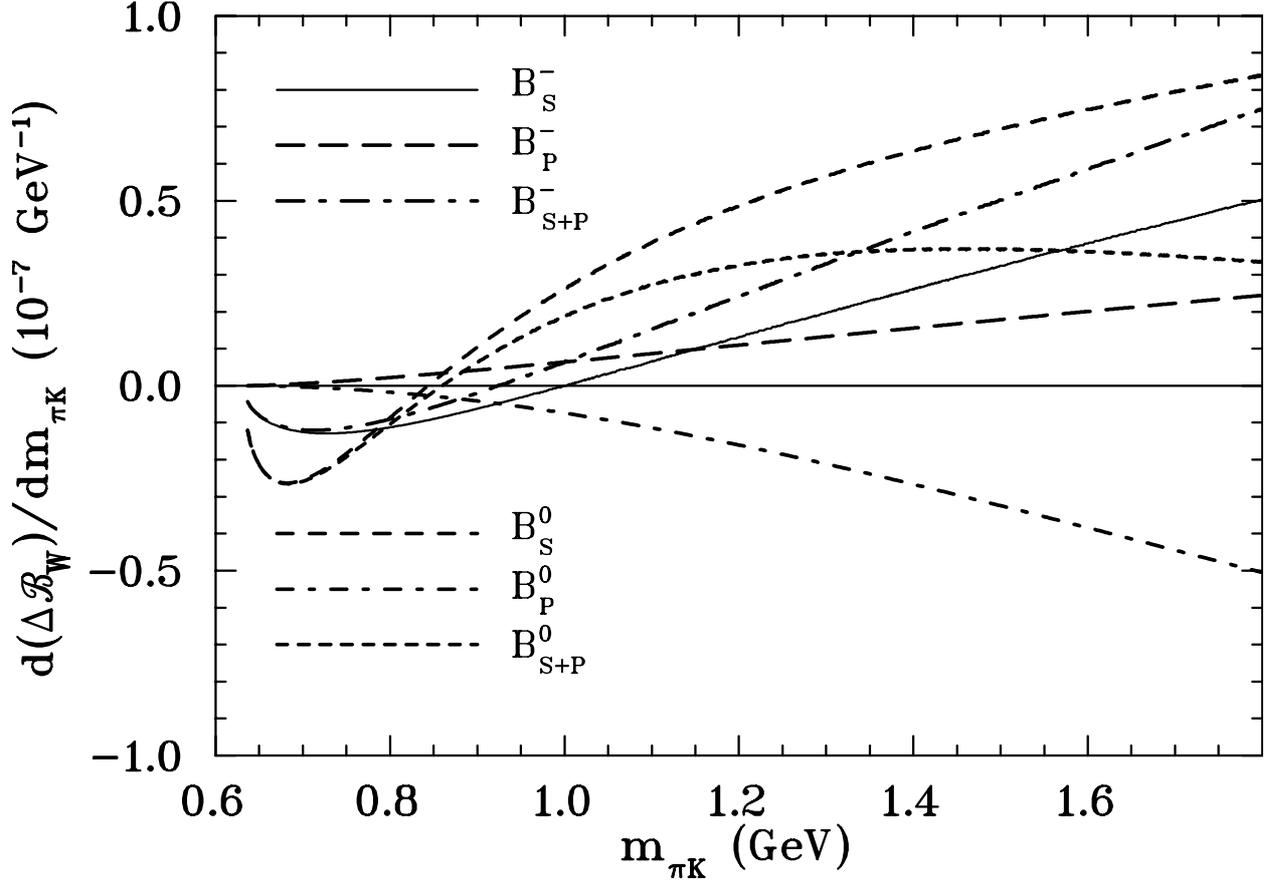}
\caption{Here $\Delta\mathcal{B}_W$ represent the contributions, to the numerator of the $CP$ 
asymmetry parameter $\mathbb{A}_{CP}$ of Eq.~(\ref{ACP}), of the different $S$, $P$, $S+P$ amplitudes where the scalar and vector form factors have been factorized out.
The curves denoted by $B^-_{S,\ P,\ S+P}$ correspond to the contribution for the  $S,\ P,\ S+P$ of this weak interaction plus perturbative QCD interaction amplitudes to the charged $B$ decays and those denoted  by $B^0_{S,\ P,\ S+P}$ the contributions to the neutral $B$ decays.\label{fig1}}
\end{figure}

\begin{figure}[ht]
\includegraphics*[angle=90,width=1.\columnwidth]{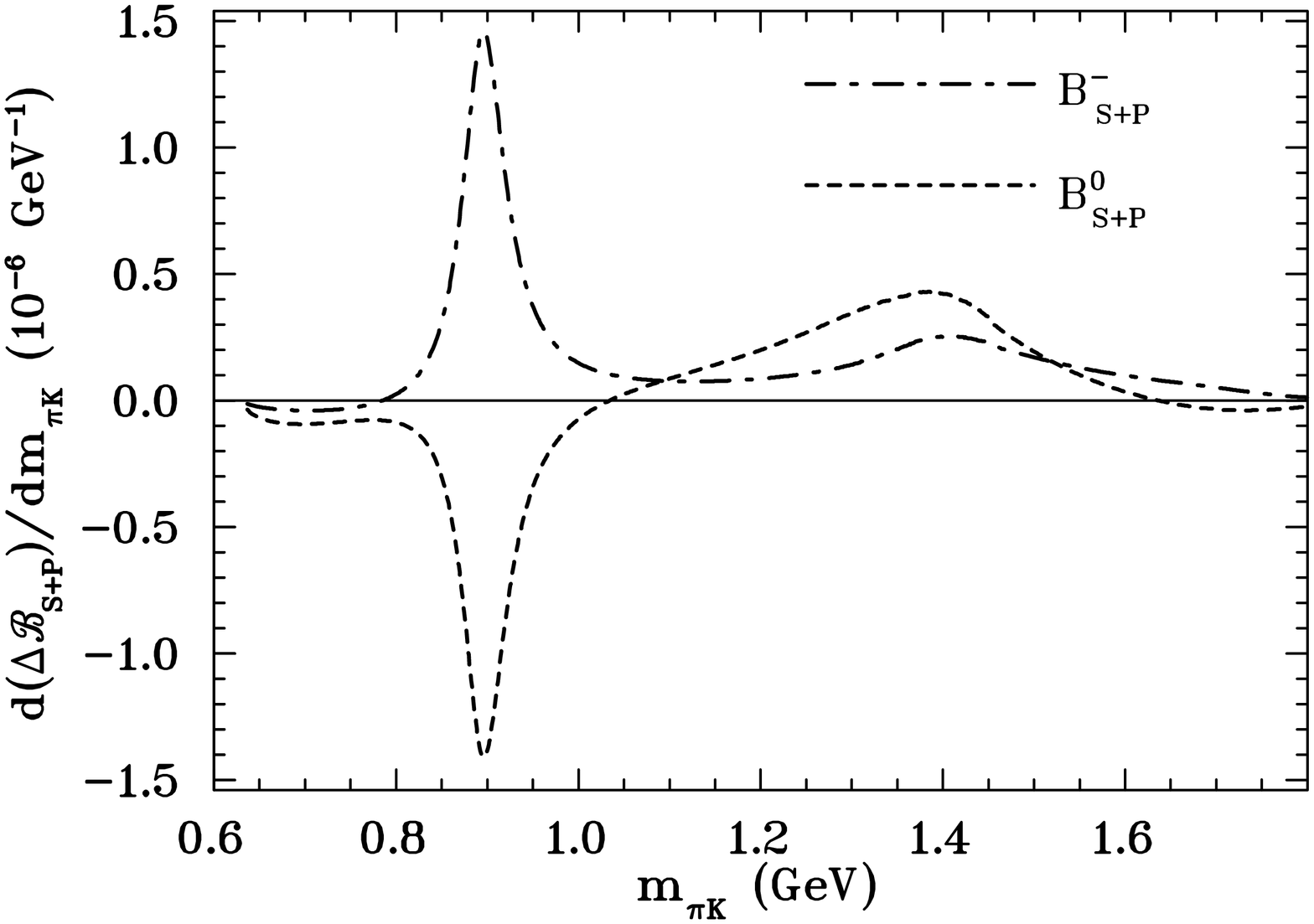}
\caption{As in Fig.~\ref{fig1} but only for the $S+P$ amplitudes including the scalar and vector form factor contributions.\label{fig2}}
\end{figure}

Within the QCDF approach including final state interactions, before and after hadronization, we fit, with the two complex parameters $(\rho_A,\ \phi_A)$ and $(\rho_H,\ \phi_H)$ the mass and helicity angle distributions, the $P$-wave branching ratios and the $CP$ asymmetries provided by the Belle~\cite{Belle75a,Belle96b,Belle05c,Belle05d} and BABAR~\cite{BaBar78a,BaBar78b,BaBar73c,BaBar07d} Collaborations.
We consider 206 effective mass distribution data, 82 helicity distribution points, 6 values of asymmetries for both $\pi K^*(892)$ and $\pi K^*_0(1430)$ and 4  branching ratios for $\pi K^*(892)$.
Altogether we have  298 observables with equal weight.
Note that in the fit  we did suppress some points which lie outside the general trend of the data.
We have checked that these suppressions do not influence the results of the fit. 

We obtain a $\chi^2/{\rm dof}=492.5/(298-4)=1.68$ with the following values  $\rho_H=54.43\pm 7.32,\ \phi_H=-0.95\pm 0.10$ radians for the hard-scattering parameters and 
$\rho_A=2.51\pm 0.11,\ \phi_A=-2.98\pm 0.06$ radians for the annihilation parameters.  
The corresponding hard-scattering, $H_n(M_2)$, contributions to the short distance amplitudes, $a_n^q(\mu)$ of Eq.~(\ref{a}), are listed in Table~\ref{ai} together with the
leading order, vertex $V_n(M_2)$ and penguin $P_n^q(M_2)$ contributions for the $P$ and $S$-waves.
The resulting annihilation amplitudes, $\beta_n(\pi M_2)$, are displayed in Table~\ref{betan}.

The amplitude, $a_n^q(\mu)$, for $n=4$ to $10$ are always corrections to the $a_1^q(\mu)$. 
For $n=1$ to $8$, the modulus of the LO contribution is larger than the modulus of the vertex term, itself larger than that of the penguin. The modulus of the hard-scattering contribution is in between $25\%$ to $60\%$ of the modulus of the LO term. 
The vertex, penguin and hard-scattering 
contributions can be seen as corrections to the leading order amplitude whereas for $n=10$, $H_n(M_2)$ gives the main contribution to the very small amplitude $a_{10}^q(\mu)$. 
The moduli of the annihilation terms  (see Table~\ref{betan})  are of the order of those of the vertex or penguin for both $P$- and $S$-waves.

Since  the present work and that of  Ref.~\cite{Bppk} (see their Table VI) use the same leading and next-to-leading order parameters, the $P$-wave vertex and penguin contributions to the $a_n^q$ are quite similar. 
For the $S$-wave, there are some differences in these corrections for $a_1^q$ and  $a_4^q$.
These arise from the introduction of Gegenbauer moments up to order 3 in Ref.~\cite{Bppk}.
The moduli of the $a_1^{q}$ are  about 20\%  smaller than those of Ref.~\cite{Bppk} (see their Table I).
This reduction comes mainly from the hard-scattering contributions. 

{\renewcommand\baselinestretch{1.10}
\begin{table*}[t]
\begin{center}
\caption{Branching fractions $\mathcal{B}$ [see Eq.~(\ref{dB-})] in units of $10^{-6}$ and direct $CP$ asymmetries $\mathbb{A}_{CP}$ in \% [Eq.~(\ref{ACP})] averaged over charge conjugate reactions. The values of the model, calculated by the integration of the $m_{\pi K}$ distribution over the $m_{\pi K}$ range from $m_{\pi K}^{min}=0.82$ to $m_{\pi K}^{max}=0.97$ GeV for the $P$ wave and from 1.0 to 1.76 for the $S$ wave are compared to the corresponding Belle and BABAR results given in the fourth column for $\mathcal{B}$ and fifth column for $\mathbb{A}_{CP}$. Model uncertainties arise from the phenomenological parameter errors obtained through the minimization. The third column gives the model values without the phenomenological hard scattering and annihilation contributions. \label{Br}}
\begin{tabular}{|cccccc|}
 \hline 
       $\mathcal{B}$(decay channel)                                     &     Model               & $H_n[\beta_n]\equiv 0$ & $\mathcal{B}^{exp}(m_{\pi K}^{min},m_{\pi K}^{max})$   &  $\mathcal{B}^{exp}$   & Refs.\\
\hline \hline 
\multirow{2}{*}{ $\mathcal{B}( B^- \to \pi^- \bar{K}^{*0} \to \pi^- \pi K)$  }        & \multirow{2}{*}{$5.82\pm0.15$} &  \multirow{2}{*}{2.17}  & $5.35\pm 0.59$ & $6.45 \pm 0.71 $  &  \cite{Belle96b}   \\  
                                            &                       &                         & $5.98\pm 0.75 $& $7.20 \pm 0.90 $   & \cite{BaBar78a}   \\
\multirow{2}{*}{  $\mathcal{B}( \bar{B}^0 \to \pi^+ \bar{K}^{*-}\to \pi^+ \pi K)$ }  &  \multirow{2}{*}{$4.50\pm 0.21$}&  \multirow{2}{*}{1.65}  & $4.65\pm 0.77$ & $5.60 \pm 0.93 $ &  \cite{Belle75a}   \\
                                            &                       &                         & $6.47 \pm 0.72$ & $11.70 \pm 1.30$   & \cite{BaBar07d} \\
\hline  
\multirow{2}{*}{  $\mathcal{B}( B^- \to \pi^- \bar{K}^{*0}_0\to \pi^- \pi K)$ }      & \multirow{2}{*}{$12.11\pm 0.32$}&  \multirow{2}{*}{7.80}  & $25.92\pm 2.45$ & $32 \pm 3.02  $  &   \cite{Belle96b}  \\
                                            &                       &                         & $17.64 \pm 3.60$ & $24.5 \pm 5.0$    &  \cite{BaBar78a}\\
\multirow{2}{*}{  $\mathcal{B}( \bar{B}^0 \to \pi^+ \bar{K}^{*-}_0\to \pi^+ \pi K)$} & \multirow{2}{*}{ $11.05\pm 0.25$ }& \multirow{2}{*}{7.45}   & $24.95\pm 3.25$ & $30.80 \pm 4.01 $  &  \cite{Belle75a} \\
                                            &                       &                         & $12.19 \pm 3.26$ & $25.40 \pm 6.80$   &  \cite{BaBar78b}  \\ 
\hline  
 \hline 
       $\mathbb{A}_{CP}$(decay channel)                                     &     Model               & $H_n[\beta_n](\pi M_2)\equiv 0$ &    &  $\mathbb{A}_{CP}^{exp}$   & Refs.\\
\hline \hline 
\multirow{2}{*}{  $\mathbb{A}_{CP}( B^- \to \pi^- \bar{K}^{*0}\to \pi^- \pi K)$   }   & \multirow{2}{*}{$0.89\pm 0.23$}                      &  \multirow{2}{*}{1.29} &  &  $-14.90 \pm 6.75 $ &    \cite{Belle96b}      \\
                                             &                       &              &  &  $3.2 \pm 5.4     $   & \cite{BaBar78a} \\
  $\mathbb{A}_{CP}(\bar{B}^0 \to \pi^+ \bar{K}^{*-}\to \pi^+ \pi K)$ &  $-0.99 \pm  3.42$                 &        7.99       &  &  $-14 \pm 12      $ &  \cite{BaBar78b}  \\
 \hline 
\multirow{2}{*}{$\mathbb{A}_{CP}( B^- \to \pi^- \bar{K}^{*0}_0\to \pi^- \pi K$) }     & \multirow{2}{*}{$0.27\pm 0.10$}                   &  \multirow{2}{*}{0.27}         & &  $7.60 \pm 4.66$        &   \cite{Belle96b}    \\
                                             &                       &              & &  $3.20 \pm 4.60$        &   \cite{BaBar78a}  \\
$\mathbb{A}_{CP}( \bar{B}^0 \to \pi^+ \bar{K}^{*-}_0\to \pi^+ \pi K)$&   $0.75\pm 0.90$                   &   -0.68           &  & $17.0 \pm 26$    &    \cite{BaBar78b}      \\
\hline  
\end{tabular}
\end{center}

\end{table*}}

The $K^{\pm} \pi^{\mp}$ effective mass distributions for $B^0 \to \pi^- \pi^+ K^0$ are globally well fitted: for  $\bar{B}^0$ decay, 
$\chi^2_{\rm Belle}/{\rm dof}=1.03$, $\chi^2_{\rm BABAR}/{\rm dof}=0.71$ and for $B^0$ decay, $\chi^2_{\rm Belle}/{\rm dof}=1.0$, $\chi^2_{\rm BABAR}/{\rm dof}=2.96$. 
In the case of the $B^{\pm} \to \pi^{\pm} \pi^{\mp} K^{\pm}$ effective mass distributions, one has $\chi^2_{\rm Belle}/{\rm dof}=2.55$,  $\chi^2_{\rm BABAR}/{\rm dof}=2.65$, 
the data being not very well reproduced, in particular for the charged $B$ decays, below 0.9 GeV. 
The helicity angle distributions are  well fitted for both decays 
with a $\chi^2/{\rm dof}$ of the order of 1. 

All the results on branching ratios and asymmetries are 
summarized in Table~\ref{Br}.
For the $K^*(892)$ branching ratios, $90\%$ of the $\chi^2/{\rm dof}$ comes from the $B^0$ and $\bar{B}^0$ BABAR 
data.
These are incompatible with the corresponding ones from Belle. The $P$-wave experimental branching ratios 
for $B^{\pm} \to \pi^{\pm} \pi^{\mp} K^{\pm}$ are well reproduced whereas our predictions for the $S$-wave branching ratios do not fully agree with 
those provided by Belle but do agree better with the  BABAR data.
As discussed in details in Ref.~\cite{Bppk}, the determination of the $B \to \pi K_0^{*} (1430)$ branching ratios is problematic as the $K_0^{*} (1430)$  resonance is wide and the result is quite model dependent.
However, within the factorization and quasi two-body hypotheses, the use of a scalar form factor, determined with precision from theory and experiments other than those of $B$ decays, makes our $\pi K_0^{*} (1430)$ branching ratio predictions well founded.

It is difficult to draw any firm
 conclusions  from the small asymmetries obtained from our 
global fit for both $P$ and $S$ waves since the experimental data have large uncertainties.  
We found that, if we introduced some factor in the $\chi^2$ to increase the weight of the $CP$ asymmetries, as done in Ref.~\cite{Bppk}, we obtain a fit of equivalent quality with, indeed, $\mathbb{A}_{CP}$ values  closer to the central  values  of the experimental analyzes, in particular for neutral $B$ decays. 

The plots on effective mass and helicity angle distributions,
almost identical to those published in~\cite{Bppk}, will not be given here. 
For the $S$-wave and for $m_{\pi K} \lesssim 0.8$~GeV, the effective mass distributions, mainly for the charged $B$ decays, are smaller than those  of Ref.~\cite{Bppk} which could indicate some stronger suppression of the $K^*_0(800)$ contribution.

In relation with the direct $CP$ violation asymmetries, we will
focus on the differential difference of effective mass branching ratio distributions for charge conjugate channels.
In Fig.~\ref{fig1} we draw $d(\Delta \mathcal{B})/dm_{\pi K}$ with $\Delta \mathcal{B}=\mathcal{B}(B \to \pi \pi K)-\mathcal{B}(\bar{B} \to \bar{\pi} \bar{\pi} \bar{K})$ [numerator of  $\mathbb{A}_{CP}$, see Eq.~(\ref{ACP})] for the charged and neutral decays and calculated from the $S$, $P$, and $S+P$ amplitudes, where the strong interaction scalar and vector form factors have been factorized out.
Figure~\ref{fig2} illustrates these distribution differences for the full $S+P$ amplitude including these form factors.
The weak interaction plus the  strong interaction before hadronization produces $S+P$  distribution differences (see Fig.~\ref{fig1}) negative for $m_{\pi K}$ below $\sim$1~GeV, positive and increasing above.
Including the final state interaction after hadronization  the $S+P$  distributions, as seen in Fig.~\ref{fig2} are enhanced in the vicinity of the  $K^*(892)$ resonance, that of the charged channel is positive while that of the neutral is negative. The positive enhancement at the 1430 resonance for the $B^0$ decays 
is larger than that of the $B^-$.

The denominator of $\mathbb{A}_{CP}$ giving similar contribution for charged and neutral channels, the above behavior of the $S+P$  distributions  allows us to understand the model values (calculated by integrating distributions over the $m_{\pi K}$ range quoted in Table~\ref{Br} caption) for $\mathbb{A}_{CP}$ displayed in Table~\ref{Br}, knowing that  the $P$-wave contribution dominates in the vector resonance region and the $S$-wave in the scalar one.
One can see that a strong final state interaction after hadronization can increase the $CP$ asymmetry.

\section{Summary and Outlook}\label{Conclusion}

In the present study, we analyze the $K^*$ resonance effects on the direct $CP$ violation in the $B\to \pi \pi K$ decay channels.
We calculate the amplitudes for the $B^0 \to \pi^- \pi^+ K^0$ and $B^{\pm} \to \pi^{\pm} \pi^{\mp} K^{\pm}$  decays in the QCD factorization framework~\cite{QCDF,Beneke:2001ev}  at leading order in $\Lambda_{QCD}/m_b$  and at the next-to-leading order in $\alpha_s$.
In order to do so, we approximate these three-body processes  as quasi two-body $B$ decays into $\pi K^*(892)$ and $\pi K^*_0(1430)$ since these final state $K^*$ resonances dominate the $\pi K$ effective mass region below 2 GeV. 
All the contributions, before hadronization, i.e., from vertex, penguin, hard-scattering and annihilation corrections as well as those after hadronization, i.e., from the $K^*$ meson resonance formation and decay described by the strong interaction scalar and vector form factors, are included.
We complete the calculation performed in Ref.~\cite{Bppk}  by adding explicitly the hard scattering and annihilation contributions which are however subject to large uncertainties arising from the presence of end-point divergences.
These divergences are modeled with two complex parameters;  they are the sole fitted parameters entering in the present calculation.
Thus, as compared to Ref.~\cite{Bppk}, our model involves   only 4 real phenomenological parameters instead of 8 while reproducing equally well the present data. 
These  4 parameters  are then determined through a fit to the available data on mass and helicity angle distributions, branching ratios and $CP$ asymmetries originating from Belle and BABAR Collaboration measurements.
The large experimental uncertainties in $CP$ asymmetries do not yield strong constraints.
Producing higher statistics experimental data seems to us mandatory in order to improve
constraints on models.
Furthermore, it should sort out the present discrepancies between the Belle and BABAR analyses.

At this stage one cannot conclude that the data is or is not compatible with the Standard Model.
Yet, the possibility of new physics effects, as, for instance, in the minimal supersymmetric  Standard Model approach studied in Ref.~\cite{Beneke2009}, cannot be excluded.
However, the theoretical basis of our model being restricted to next-to-leading order corrections, the phenomenological terms can simulate next-to-next-to leading order (NNLO) effects.
It could also take into account charming penguin contributions.
In principle, the NNLO corrections to hard scattering are amenable to convergent integrals which can be evaluated~\cite{Beneke2006,BenekeNNLO}.
This contribution could reduce the phenomenological part of our model amplitudes.
The long distance charming penguin amplitudes such as those arising from intermediate $D_s^{(*)} D^{(*)}$ states could be important since the branching fractions for the transition $B\to D_s^{(*)} D^{(*)}$ are quite large.
However, their contributions cannot be calculated in a QCD pertubative framework.
Both NNLO corrections and charming penguin amplitudes should be included before being able to give firm statement as to wether or not it is necessary to introduce new physics to understand the data, but, this is outside the scope of the present study.

In conclusion, from  this analysis, we point out the important following aspects.

$\bullet$ It constitutes a robust state of the art QCD factorization calculation at next-to-leading order in the strong coupling constant. 
In this framework, the strong phase can be generated dynamically.
However, the mechanism suffers from end-point singularities which are not well controlled. 
It is now apparent that the
Cabibbo-Kobayashi-Maskawa matrix is the dominant source of CP
violation in flavour changing processes in B decays. The corrections
to this dominant source coming from beyond the Standard Model are not
expected to be large. 
In fact, the main remaining uncertainty
lies in the factorization approximation which provides an explicit picture  in the
heavy quark limit. It takes into account all the leading contributions
as well as subleading corrections to the na\"ive factorization. 
The soft collinear effective theory (SCET) has been proposed as a new procedure
for factorization~\cite{Beneke2006}. It allows one to formulate  a
collinear factorization theorem in terms of effective operators where
new effective degrees of freedom are involved, in order to take into
account the collinear, soft, and ultrasoft quarks and gluons.
Following such steps should  improve further our knowledge
of B physics and, eventually, hint at contributions from physics
beyond the Standard model. 

$\bullet$ It illustrates explicitly  how the strong final state
interaction after hadronization can enhance $CP$ violation asymmetries.
The variation of the differential difference of effective mass branching ratio distribution for charge conjugate channels as a function of the $\pi K$ invariant mass over
the whole range of the $K^*(892)$ and $K^*_0(1430)$  resonances shows that mixing
resonance effects, as those seen in Fig.~\ref{fig2}, 
can be observed within a window of 100-200 MeV.
With the new Large Hadron Collider (LHC) providing energy and
accuracy (small energy bin), we believe that by exploring such
windows the LHCb Collaboration should be in a
position to perfom accurate measurements of CP violation in $B$
to $\pi \pi K$ decays. 

$\bullet$ It confirms the advantage of using, as a consequence
of  QCD factorization, a scalar form factor to
describe the $\pi K_0^*(1430)$ final state. 
The  $K_0^*(1430)$ resonance is very wide and  its nonresonant part is difficult to evaluate. 
Thus, the determination of the $B \to \pi K^*_0(1430)$ branching fractions within, in particular, the isobar model, leads to large uncertainties.
As advocated in Ref.~\cite{Bppk}, a parametrization with this scalar form factor, precisely constructed from unitary coupled channel equations using experimental kaon-pion $T$-matrix elements  together with chiral symmetry and asymptotic QCD constraints, should be used in experimental Dalitz plot analysis.

\begin{acknowledgments}

We thank the non-participating authors of Ref.~\cite{Bppk} for their kind support during the course of this work. This research has been financed in part by an {\em IN2P3-CNRS\/} 
theory grant for the project ``{\em Contraintes sur les phases fortes dans les d\'esint\'egrations hadroniques des m\'esons B\/}" and by the IN2P3-Polish Laboratories Convention (Project No 08-127).
\end{acknowledgments}


\begin{thebibliography}{00}

\bibitem{CKM}
  M.~Kobayashi and T.~Maskawa,
  Prog.\ Theor.\ Phys.\   {\bf 49}, 652 (1973), \textit{$CP$-Violation in the Renormalizable Theory of Weak Interaction}; 
 N.~Cabibbo,
  Phys.\ Rev.\ Lett.\   {\bf 10}, 531 (1963), \textit{Unitary Symmetry and Leptonic Decays}.
\bibitem{QCDF}
  M.~Beneke and M.~Neubert,
 Nucl. Phys. {\bf B675}, 333 (2003),  \textit{QCD factorization for $B\to PP$ and $B\to PV$ decays}.
\bibitem{Bppk}
 B. El-Bennich, A.~Furman, R.~Kami\'nski, L.~Le\'sniak,  B.~Loiseau, B.~Moussallam,  Phys. Rev. D {\bf 79}, 094005 (2009), \textit{$CP$ violation and kaon-pion interactions in $B \to K \pi^+ \pi^- $ decays}.
\bibitem{Moussallam:2007qc}
B.~Moussallam, Eur. Phys. J. C {\bf 53}, 401 (2008), \textit{Analyticity constraints on the strangeness changing vector current and applications to $\tau\to K\pi \nu_\tau$, $\tau\to K\pi\pi \nu_\tau$}.

\bibitem{PDG}
 C. Amsler {\it et al.} (Particle Data Group),
  Phys. Lett. B {\bf 667},1 (2008), \textit{Review of particle physics}.
\bibitem{Beneke:2001ev}
  M.~Beneke, G.~Buchalla, M.~Neubert and C.T.~Sachrajda, Nucl.\ Phys.\  {\bf B606}, 245 (2001),  \textit{QCD factorization in $B \to \pi K, \pi \pi$ decays and extraction of Wolfenstein  parameters}.
 
\bibitem{Cheng:2005nb}
  H.~Y.~Cheng, C.K.~Chua and K.-C. Yang, Phys.\ Rev.\  D {\bf 73}, 014017(2006), \textit{Charmless hadronic $B$ decays involving scalar mesons: Implications to the nature of light scalar mesons }.

\bibitem{Melikhov:2001zv}
  D.~Melikhov,
  Eur.\ Phys.\ J.\ direct {\bf C2}, 1 (2002), {\it Dispersion approach to quark-binding effects in weak decays of heavy mesons}.


\bibitem{Belle75a}
  A.~Garmash  {\it et al.} (Belle Collaboration), Phys.\ Rev.\  D  {\bf 75}, 012006 (2007){, \it Dalitz analysis of three-body charmless $B^0 \to  K^0 \pi^+ \pi^-$ decay}.

\bibitem{Belle96b}
 A.~Garmash {\it et al.}  (Belle Collaboration), Phys.\ Rev.\ Lett.\   {\bf 96}, 251803 (2006), {\it Evidence for large direct $CP$ violation in $B^\pm \to \rho(770)^0 K^\pm$ from analysis of the three-body charmless $B^\pm \to K^\pm \pi^\pm \pi^\mp$ decay}.


\bibitem{Belle05c}
 K.~Abe {\it et al.} (Belle Collaboration), arXiv: hep-ex/0509001, \textit{Search for Direct CP Violation in Three-Body Charmless $B^\pm \to  K^\pm \pi^\pm \pi^\mp$ Decay}.

\bibitem{Belle05d}
K.~Abe  {\it et al.} (Belle Collaboration), arXiv: hep-ex/0509047,{\it  Dalitz analysis of the three-body charmless decay $B^{0} \to K^0_S \pi^+ \pi^-$}. 

\bibitem{BaBar78a}
  B.~Aubert  {\it et al.} ($BABAR$ Collaboration), Phys.\ Rev.\  D  {\bf 78}, 012004(2008), \textit{ Evidence for Direct CP Violation from Dalitz-plot analysis of $B^\pm \to  K^\pm \pi^\pm \pi\mp$}. 

\bibitem{BaBar78b}
 B.~Aubert {\it et al.}  ($BABAR$ Collaboration), Phys.\ Rev.\  D {\bf 78},   052005 (2008), \textit{ Dalitz Plot Analysis of the Decay $B^0(\bar B^0) \to K^\pm \pi\mp \pi^0$}.

\bibitem{BaBar73c}
B.~Aubert  {\it et al.} ($BABAR$ Collaboration), Phys.\ Rev.\  D {\bf 73}, 031101 (2006), {\it Measurements of neutral $B$ decay branching fractions to  $K^0_S \pi^+ \pi^-$ final states and the charge asymmetry of $B^{0} \to K^{*+} \pi^-$}.


\bibitem{BaBar07d}
 B.~Aubert  {\it et al.} ($BABAR$ Collaboration), arXiv: 0708.2097 [hep-ex], \textit{Time-dependent Dalitz Plot Analysis of $B^0 \to  K_S \pi^+ \pi^-$}.

\bibitem{Beneke2009}
M. Beneke, Xin-Qiang Li, L. Vernazza, 
 Eur. Phys. J. C {\bf 61}, 429 (2009), \textit{Hadronic $B$ decays in the MSSM with large $\tan \beta$}.

\bibitem{Beneke2006}
M. Beneke, Nucl. Phys. B (Proc. Suppl.) \textbf{170}, 57 (2007)
, \textit{Hadronic B decays}.


\bibitem{BenekeNNLO}
M. Beneke, T. Huber, Xin-Qiang Li,
Nucl. Phys. {\bf B832}, 109 (2010), \textit{NNLO vertex corrections to non-leptonic B decays: Tree amplitudes}.

 
\end{thebibliography}
\end{document}